\documentclass{article}
\usepackage{amsthm}
\usepackage{amsmath}
\usepackage{amssymb}
\usepackage{graphicx}
\usepackage{subcaption}
\usepackage{enumitem}
\usepackage{tikz}
\usepackage{physics}
\usepackage{braket}
\usepackage{mathtools}
\usepackage{algorithm}
\usepackage{algcompatible}

\usepackage{hyperref}
\usepackage{authblk}
\usepackage{booktabs}
\usepackage{fourier}
\usepackage[numbers]{natbib}

\newtheorem{theorem}{Theorem}

\newcommand{\acks}[1]{\section*{Acknowledgments}#1}

\title{Faster Stochastic First-Order Method for Maximum-Likelihood \\ Quantum State Tomography}

\author[1]{Chung-En Tsai}
\author[2,3,4,5]{Hao-Chung Cheng}
\author[1,3,5]{Yen-Huan Li}
\affil[1]{Department of Computer Science and Information Engineering, National Taiwan University}
\affil[2]{Department of Electrical Engineering and Graduate Institute of Communication Engineering, National Taiwan University}
\affil[3]{Department of Mathematics, National Taiwan University}
\affil[4]{Hon Hai (Foxconn) Quantum Computing Centre}
\affil[5]{Center for Quantum Science and Engineering, National Taiwan University}

\date{}

\newcommand{\N}{\mathbb{N}}
\newcommand{\D}{\mathcal{D}}
\newcommand{\C}{\mathbb{C}}

\DeclareMathOperator{\diag}{diag}
\DeclareMathOperator*{\argmin}{arg\,min}

\begin{document}

\maketitle

\begin{abstract}%
In maximum-likelihood quantum state tomography, both the sample size and dimension grow exponentially with the number of qubits. 
It is therefore desirable to develop a stochastic first-order method, just like stochastic gradient descent for modern machine learning, to compute the maximum-likelihood estimate. 
To this end, we propose an algorithm called stochastic mirror descent with the Burg entropy. 
Its expected optimization error vanishes at a $O ( \sqrt{ ( 1 / t ) d \log t } )$ rate, where $d$ and $t$ denote the dimension and number of iterations, respectively. 
Its per-iteration time complexity is $O ( d^3 )$, independent of the sample size. 
To the best of our knowledge, this is currently the computationally fastest stochastic first-order method for maximum-likelihood quantum state tomography. 
\end{abstract}

\section{Introduction} \label{sec_intro}

In maximum-likelihood quantum state tomography \citep{Hradil1997aa}, the estimate of the unknown quantum state is given by 
\begin{equation} \label{eq:1}
	\hat{\rho} \in \argmin_{\rho\in\D} f(\rho),\quad\text{where }f(\rho)\coloneqq\frac{1}{n}\sum_{i=1}^n -\log\tr(A_i\rho),
\end{equation}
where $A_i$ are Hermitian positive semi-definite matrices determined by the measurement outcomes and $\mathcal{D}$ denotes the set of density matrices, i.e., 
\[
\D\coloneqq\qty{\rho\in\C^{d\times d}:\rho^\ast=\rho,\rho\geq 0,\tr\rho=1} . 
\]
We call $n$ the \emph{sample size}. 

The optimization problem \eqref{eq:1} is apparently convex. 
Solving it rigorously and efficiently, however, is surprisingly difficult. 
The first challenge is the computational scalability with respect to the sample size $n$. 
Consider computing $\hat{\rho}$ by projected gradient descent, perhaps the simplest convex optimization algorithm \citep{Bolduc2017}. 
Projected gradient descent iterates as 
\[
\quad \rho_{t + 1} = \Pi_{\mathcal{D}} \left( \rho_t - \eta \nabla f ( \rho_t ) \right) , \quad \forall t \in \mathbb{N} , 
\]
given some initial iterate $\rho_1 \in \mathcal{D}$ and step size $\eta > 0$, where $\Pi_{\mathcal{D}}$ denotes projection onto $\mathcal{D}$ with respect to the Frobenius norm. 
In each iteration, computing the projection onto $\mathcal{D}$ takes time $O ( d^3 )$ \citep{Kyrillidis2013}; 
computing the gradient, 
\[
\nabla f ( \rho_t ) = \frac{1}{n} \sum_{i = 1}^n \frac{ - A_i }{ \tr ( A_i \rho_t ) } , 
\] 
takes time $O ( n d^\omega )$, where $\omega$, $2 \leq \omega < 2.373$ is the matrix multiplication exponent \citep{Alman2021}; 
the overall per-iteration time is hence $O ( d ^ 3 + n d ^ \omega  )$. 
It has been shown that any quantum state tomography scheme with incoherent measurements, including maximum-likelihood quantum state tomography, requires $n = \Omega ( d^3 / \varepsilon ^ 2 )$ to achieve an estimation error smaller than $\varepsilon$ in the trace norm \citep{Chen2022aa}. 
The dependence of the per-iteration time on the sample size $n$ becomes the main computational bottleneck. 

Another challenge lies in the fact that rigorously speaking, projected gradient descent is not directly applicable to solving \eqref{eq:1}. 
The reason is two-fold. 
First, the function $f$ is neither Lipschitz nor Lipschitz gradient, violating standard assumptions in the literature (see, e.g., the textbook by \citet{Nesterov2018a}). 
It is unclear how to choose the step size $\eta$ to guarantee convergence for projected gradient descent. 
Newton's method is guaranteed to converge \citep{Nesterov2018a}. 
Regarding the exponentially growing dimension, however, computations involving the Hessian are computationally too expensive. 
Second, in standard implementations of maximum-likelihood quantum state tomography, the matrices $A_i$ are low rank, either of rank $d / 2$ with Pauli measurements or of rank $1$ with Pauli basis measurements. 
Projection onto $\mathcal{D}$ typically results in low-rank matrices \citep{Kyrillidis2013}. 
The trace of the product of two low-rank matrices can be exactly zero. 
It can happen that $\tr ( A_i \rho_t ) = 0$ for some $A_i$ and some iterate $\rho_t$ of projected gradient descent; 
then, $f ( \rho_t )$ and $\nabla f ( \rho_t )$ are undefined and projected gradient descent is forced to ``stall.''
We refer to the discussions by \citet{Knee2018a} and \citet{You2022} for the details. 

Previously, we developed an algorithm, Stochastic Q-Soft-Bayes, to address the two challenges above \citep{Lin2021b}. 
\begin{itemize}
\item It only involves one randomly selected $A_i$, like stochastic gradient descent in modern machine learning, in each iteration. 
Therefore, its per-iteration time is $O ( d^3 + d^\omega )$, independent of the sample size $n$\footnote{The term $d^\omega$ is irrelevant as $\omega < 3$. We include it simply for the reader to easily check how the time complexity is evaluated.}. 
\item It always generates full-ranked iterates, thereby avoiding the ``stalling'' issue\footnote{Notice that this does not prevent the algorithm from converging to a singular minimizer.}. 
\item It has a rigorous non-asymptotic convergence guarantee. 
Its expected optimization error vanishes at a $O ( \sqrt{ ( 1 / t ) d \log d } )$ rate. 
\end{itemize}

In this paper, we propose a completely different algorithm, stochastic mirror descent with the Burg entropy. 
The algorithm shares the same three merits above, except that the convergence rate is $O ( \sqrt{ ( 1 / t ) d \log t } )$, slightly worse than Stochastic Q-Soft-Bayes when $t \gg d$. 
Nevertheless, its per-iteration time is significantly shorter. 
Each iteration of Stochastic Q-Soft-Bayes requires computing two matrix logarithms and one matrix exponential, whereas each iteration of stochastic mirror descent with the Burg entropy only requires computing one eigendecomposition. 
Though for both algorithms, the per-iteration time is $O ( d^3 )$ in theory, the difference in per-iteration time is significant in practice. 
Numerical results verify the competitiveness of the proposed algorithm compared to Stochastic Q-Soft-Bayes. 
The results also show the stochastic methods have some space for improvement compared to non-stochastic ones. 

\section{Related Work} \label{sec_related_work}
There are several approaches to quantum state tomography \citep{Nielsen2010,Hradil1997aa,Gross2010,Blume-Kohout2010,ODonnell2016,Haah2017,Torlai2018,Guta2020}. 
This paper focuses on the maximum-likelihood approach.
In particular, this papers focuses on how to compute the maximum-likelihood estimate efficiently. 
Notice that the statistical performance---the statistical estimation error--is already determined by the formulation of the maximum-likelihood estimator. 
The interested reader is referred to, e.g., \citet{Scholten2018} for a statistical analysis of the maximum-likelihood estimator. 

As discussed in Section \ref{sec_intro}, textbook convex optimization algorithms do not direct apply to computing the maximum-likelihood estimate. 
There are some advanced algorithms that are applicable, such as diluted $R \rho R$ \citep{Rehacek2007aa}, SCOPT \citep{Tran-Dinh2015aa}, NoLips \citep{Bauschke2017}, entropic mirror descent with line search \citep{Li2019a}, and several variants of the Frank-Wolfe method \citep{Dvurechensky2020,Carderera2021aa,Zhao2022aa}.
Among them, diluted $R \rho R$ and entropic mirror descent with line search are only guaranteed to converge asymptotically; 
SCOPT only has a local convergence rate guarantee. 
Hence, their computational complexities are unclear. 
NoLips and the Frank-Wolfe methods proposed by \citet{Dvurechensky2020} and \citet{Carderera2021aa} all converge at a $O ( 1 / t )$ rate; 
the dependence of their convergence rates on the problem parameters, such as the dimension and sample size, are unclear. 
Hence, it is unclear whether they scale with the dimension and sample size or not. 
The Frank-Wolfe method proposed by \citet{Zhao2022aa} converges at a $O ( n / t )$ rate, but each iteration of the algorithm involves Hessian computations, computationally expensive under the quantum setup. 

For computing a quantum state estimate by stochastic first-order methods, we are only aware of two existing results. 
\citet{Youssry2019aa} proposed stochastic entropic mirror descent and proved its asymptotic convergence for the least-squares regression approach to quantum state tomography. 
\citet{Lin2021b} proposed Stochastic Q-Soft-Bayes for the maximum-likelihood approach and derived a non-asymptotic convergence guarantee, as discussed in Section \ref{sec_intro}. 
This paper aims to improve on the algorithm by \citet{Lin2021b} in terms of the computational efficiency. 

There are a few online learning formulations of estimating quantum states. 
\citet{Yang2020} studied several online convex optimization algorithms for Lipschitz losses. 
Unfortunately, as mentioned in Section \ref{sec_intro}, the logarithmic loss in \eqref{eq:1} is not Lipschitz. 
\citet{Zimmert2022} developed an algorithm for online learning quantum states with the logarithmic loss. 
Their algorithm, after an online-to-batch conversion (see the discussion in Section \ref{sec_alg}), also solves \eqref{eq:1}.
The algorithm, however, involves $t - 1$ data points in the $t$-th iteration and adopts Newton's method to compute each iterate. 
Therefore, the algorithm does not scale with the sample size and dimension. 

A closely related but essentially different problem is shadow tomography \citep{Aaronson2020}. 
Shadow tomography aims to estimate the probability distribution of measurement outcomes instead of the whole density matrix. 
Therefore, it has been proven that shadow tomography requires a much smaller sample size than state tomography. 
Shadow tomography can be implemented by an online learning algorithm \citep{Aaronson2018}. 
Though shadow tomography is efficient in sample size, we notice that the online shadow tomography algorithm still has per-iteration time $O ( d^3 )$ as our algorithm. 

\section{Algorithm and Convergence Guarantee}

\subsection{Stochastic mirror descent with the Burg entropy} \label{sec_alg}

The proposed algorithm, stochastic mirror descent with the Burg entropy, is shown in Algorithm \ref{alg:1}. 
In the algorithm,
\[
\nabla f_t ( \overline{\rho}_t ) = \frac{ - A_{i_t} }{ \tr ( A_{i_t} \overline{\rho}_t ) } , 
\]
and $D_h$ denotes the Bregman divergence with respect to the Burg entropy, which we denote by $h$, i.e., 
\begin{align*}
& D_h ( \rho, \sigma ) \coloneqq h ( \rho ) - h ( \sigma ) - \tr ( \nabla h ( \sigma ) ( \rho - \sigma ) ) , \\
& h ( \rho ) \coloneqq - \log \det \rho , 
\end{align*}
where $\nabla h (\rho) = - \rho^{-1}$. 

\begin{algorithm}[ht] 
\caption{Stochastic mirror descent with the Burg entropy.} 
\label{alg:1}
\hspace*{\algorithmicindent} \textbf{Input: } $\eta > 0$.
\begin{algorithmic}[1]
\STATE $\rho_1 = I / d$. 
\FORALL{$t \in \mathbb{N}$}
	\STATE $\overline{\rho}_{t} \leftarrow (1/t)\sum_{\tau = 1}^t \rho_\tau$.
	\STATE Output $\overline{\rho}_t$.
	\STATE Sample an index $i_t \in \set{ 1, \ldots, n }$ uniformly and independently of the past.
	\STATE $f_t(\rho)\coloneqq -\log \tr ( A_{i_t} \rho )$.
	\STATE $\rho_{t + 1} \leftarrow \argmin_{\rho \in \mathcal{D}} \eta \tr ( \nabla f_t ( \overline{\rho}_t ) ( \rho - \rho_t ) ) + D_h ( \rho, \rho_t )$.
\ENDFOR
\end{algorithmic}
\end{algorithm}

The proposed algorithm is based on online mirror descent with the Burg entropy, a recent algorithm we developed for online learning a quantum state with the logarithmic loss. 
By the logarithmic loss, we mean loss functions of the form $\rho \mapsto - \log \tr ( A \rho )$ for some Hermitian positive semi-definite matrix $A$. 
After an online-to-batch conversion, online mirror descent with the Burg entropy can be transformed to an algorithm minimizing an expectation of the logarithmic loss \citep{Cesa-Bianchi2004aa,Cutkosky2019aa}. 
Notice that the function $f$ we want to minimize is exactly the expected logarithmic loss with respect to the uniform distribution on $\set{ A_1, \ldots, A_n }$, or the empirical distribution on the data, so online mirror descent with the Burg entropy followed by an online-to-batch conversion readily applies.  
There are several online-to-batch methods. 
Algorithm \ref{alg:1} adopts the ``anytime online-to-batch'' procedure by \citet{Cutkosky2019aa}, as its empirical performance seems to be better. 

Notice that each iteration of Algorithm \ref{alg:1} only involves one randomly selected $A_{i_t}$, instead of $A_1, \ldots, A_n$. 
Therefore, its per-iteration time is independent of the sample size $n$. 

\subsection{Efficient Implementation}

The seventh line in Algorithm \ref{alg:1} does not have a closed-form expression, but can be efficiently computed. 
Following the discussion by \citet{Kotlowski2019}, it suffices to: 
\begin{enumerate}
	\item compute the eigendecomposition of $\eta \nabla f_t ( \overline{\rho}_t ) + \rho_t^{-1}$ and get a vector $\lambda$ of eigenvalues and a matrix $U$ consisting of eigenvectors; 
	\item project the vector $\lambda$ onto the the probability simplex, the set of entrywise-nonnegative vectors of unit $1$-norm, with respect to the logarithmic barrier and get a vector $\lambda'$; and 
	\item compute the matrix $U \diag ( \lambda' ) U^*$ as $\rho_{t + 1}$. 
\end{enumerate}
The projection with respect to the logarithmic barrier can be computed by Newton's method on the real line $\mathbb{R}$, which takes very short time in practice. 
We detail the steps above in Algorithm \ref{alg:2} for the reader's convenience.
In Algorithm \ref{alg:2}, $\texttt{eig}(A)$ is a procedure that computes the eigendecomposition of $A$. 
The eigenvectors are returned in a matrix $U$.
The parameter $\varepsilon$ specifies the desired error achieved by Newton's method. 

\begin{algorithm}[t] 
\caption{Efficient computation of $\rho_{t + 1}$. }
\label{alg:2}
\hspace*{\algorithmicindent} \textbf{Input: } $\eta > 0,\varepsilon>0, A_{i_t}, \overline{\rho}_t, \rho_t, \rho_t^{-1}$.
\begin{algorithmic}[1]
\STATE $\nabla \leftarrow -A_{i_t}/\tr ( A_{i_t} \overline{\rho}_t )$.
\STATE $(\lambda_1,\ldots,\lambda_d), U \leftarrow \texttt{eig}(\eta\nabla + \rho_t^{-1})$.
\STATE $\phi(\theta) \coloneqq \theta - \sum_{i=1}^d \log(\theta + \lambda_i)$.
\STATE $\theta \leftarrow 1 - \min_{i\in[d]}\lambda_i$.
\WHILE {$\abs{\phi'(\theta)} / \sqrt{\phi''(\theta)}\geq\varepsilon$}
	\STATE $\theta \leftarrow \theta - \phi'(\theta) / \phi''(\theta)$.
\ENDWHILE
\FORALL{$i\in[d]$}
	\STATE $\lambda_i' \leftarrow (\theta + \lambda_i)^{-1}$.
\ENDFOR
\STATE $\Lambda' \leftarrow \diag (\lambda_1',\ldots,\lambda_d')$
\STATE $\rho_{t+1} \leftarrow U \Lambda U^\ast$.
\STATE $\rho_{t+1}^{-1} \leftarrow U \Lambda^{-1} U^\ast$.
\STATE \textbf{return} $(\rho_{t+1}, \rho_{t+1}^{-1})$.
\end{algorithmic}
\end{algorithm}

The computational bottleneck in Algorithm \ref{alg:2} is the eigendecomposition. 
It is easily checked, then, that the per-iteration time of Algorithm \ref{alg:1} is $O ( d^3 )$. 

\subsection{Convergence Guarantee}

The following non-asymptotic convergence guarantee of Algorithm~\ref{alg:1} follows directly from the regret bound of online mirror descent with the Burg entropy in \citep{Tsai2022aa} and the analysis of anytime online-to-batch conversion \citep{Cutkosky2019aa}. 

\begin{theorem} \label{thm:1}
	Fix $T\in\N$. Let $\qty{\overline{\rho}_t}_{t\in\N}$ be the iterates generated by Algorithm~\ref{alg:1} with
\begin{equation*}
	\eta = \frac{\sqrt{d\log T}}{\sqrt{T} + \sqrt{d\log T}}.
\end{equation*}
Then it holds that
	\begin{equation*}
		\mathsf{E}\qty[f(\overline{\rho}_T) - \min_{\rho\in\mathcal{D}}f(\rho)] \leq 2\sqrt{\frac{d\log T}{T}} + \frac{d\log T}{T} , 
	\end{equation*}
where the expectation is taken with respect to the random choices of $A_{i_t}$ in Algorithm \ref{alg:1}. 
\end{theorem}

\section{Numerical Results}

This section compares the performances of Algorithm~\ref{alg:1}, stochastic Q-Soft-Bayes, monotonous Frank-Wolfe, and $R\rho R$ \citep{Lvovsky2004aa,MolinaTerriza2004} on synthetic data. 
Stochastic Q-Soft-Bayes has been discussed in previous sections. 
Monotonous Frank-Wolfe and $R \rho R$ are not stochastic methods; 
each iteration of the algorithms require processing the whole data-set.
We include the two algorithms for comparison as they are easy to implement. 
Monotonous Frank-Wolfe is one of the variants of Frank-Wolfe discussed in Section \ref{sec_related_work}. 
Its convergence rate has unclear dependence on the dimension and sample size. 
$R \rho R$ is a predecessor of diluted $R \rho R$, discussed in Section \ref{sec_related_work}. 
$R \rho R$ typically converges fast in practice, but does not work on carefully designed data-sets. 
Notice the comparison with monotonous Frank-Wolfe and $R \rho R$ is rigorously speaking unfair, as the two algorithms lack clear non-asymptotic convergence guarantees. 

\begin{figure}[t]
\centering
\begin{subfigure}{.48\textwidth}
	\centering
	\includegraphics[width=\textwidth]{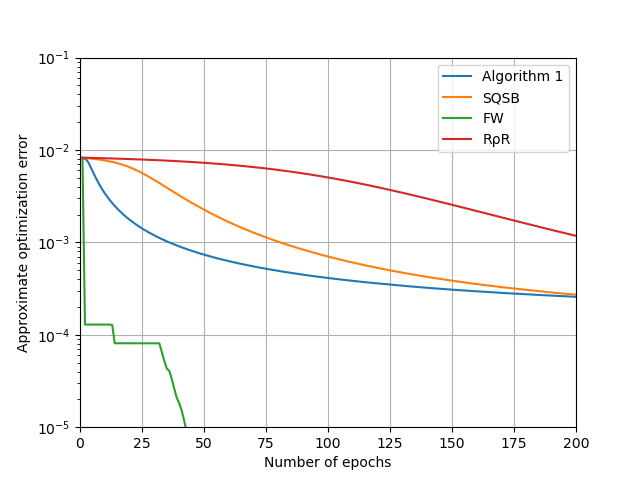}
	\caption{Approximate optimization error in function value versus the number of epochs.}
	\label{fig:1a}
\end{subfigure}%
\hfill
\begin{subfigure}{.48\textwidth}
	\centering
	\includegraphics[width=\textwidth]{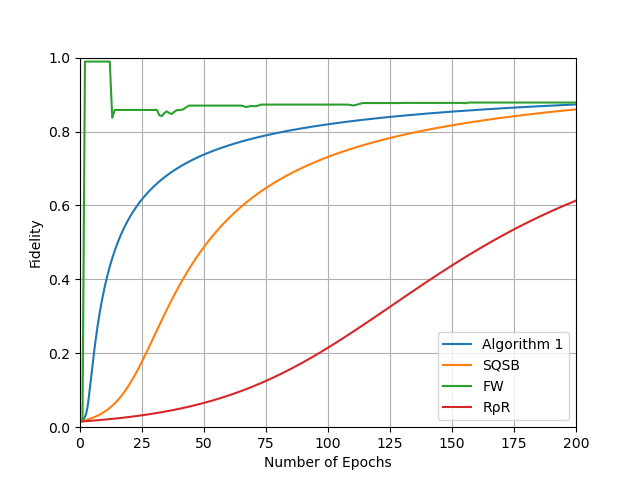}
	\caption{Fidelity between the iterates and the $W$ state versus the number of epochs.}
	\label{fig:1b}
\end{subfigure}

\begin{subfigure}{.48\textwidth}
	\centering
	\includegraphics[width=\textwidth]{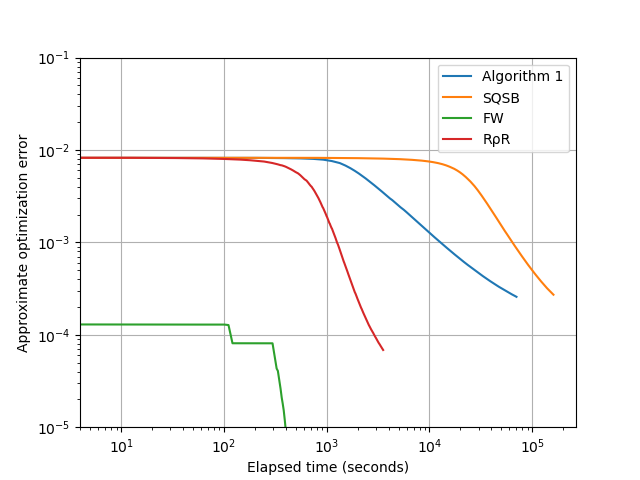}
	\caption{Approximate optimization error in function value versus the elapsed time.}
	\label{fig:1c}
\end{subfigure}%
\hfill
\begin{subfigure}{.48\textwidth}
	\centering
	\includegraphics[width=\textwidth]{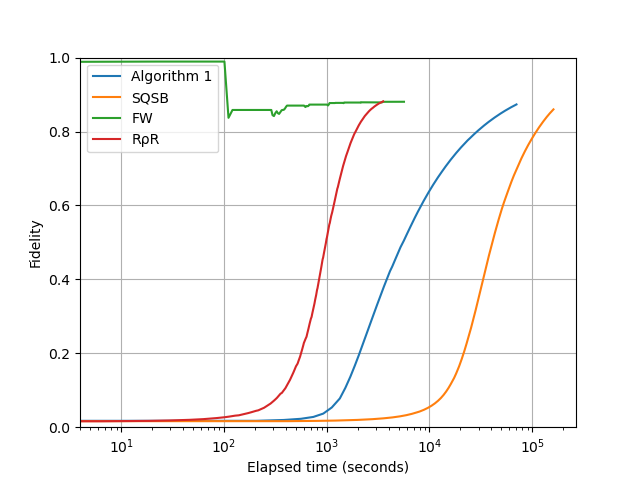}
	\caption{Fidelity between the iterates and the $W$ state versus the elapsed time.}
	\label{fig:1d}
\end{subfigure}
\caption{Comparison of the performances of Algorithm~\ref{alg:1}, stochastic Q-Soft-Bayes (SQSB), monotonous Frank-Wolfe (FW), and $R\rho R$.}
\label{fig:1}
\end{figure}

The numerical results are shown in Figure \ref{fig:1}. 
The data-set is synthetic. 
There are six qubits, so $d = 2^6 = 64$. 
The unknown density matrix to be estimated is the W-state. 
The sample size $n$ is set to $4^6 \times 100 = 409600$. 
Each data point corresponds to the measurement outcome of a randomly chosen Pauli observable. 

In Figure \ref{fig:1}, one epoch means one full pass of the data-set. 
One iteration of Algorithm \ref{alg:1} and Stochastic Q-Soft-Bayes corresponds to $1 / n$ epoch, whereas one iteration of $R \rho R$ and monotonous Frank-Wolfe corresponds to one epoch. 
The exact minimizer in \eqref{eq:1}, i.e., $\hat{\rho}$, is not accessible. 
We run all algorithms for 200 epochs and set the smallest values of $f$ achieved as $\widehat{f^\star}$. 
Then, the approximate optimization error is the difference between the value of $f$ achieved at an iterate and $\widehat{f^\star}$. 
The elapsed time is recorded on a machine with an Intel Xeon Gold 5218 CPU of 2.30GHz and 131621512kB memory. 
The four algorithms are implemented with the Julia language with the Intel Math Kernel Library (MKL). 
The number of threads for BLAS is set to $8$. 
It should be mentioned that the empirical speed depends highly on the implementations.

Regarding the number of epochs, Algorithm~\ref{alg:1} converges slightly faster than Stochastic Q-Soft-Bayes and much faster than $R\rho R$. Monotonous Frank-Wolfe is significantly faster than the others. 
An explanation given by \citet{Lin2021b} is that Frank-Wolfe prefers low-rank iterates, which matches the structure of the W-state, a rank-$1$ density matrix. Algorithm~\ref{alg:1} converges faster than Stochastic Q-Soft-Bayes in the beginning but the gap vanishes in the end. 
This may be explained by the fact that the expected optimization error of Algorithm~\ref{alg:1} is $O(\sqrt{d\log T}/\sqrt{T})$, which is slightly worse than $O(\sqrt{d\log d}/\sqrt{T})$ of stochastic Q-Soft-Bayes when $T \gg d$.

Regarding the elapsed time, Algorithm~\ref{alg:1} is about $2.3$-times faster than Stochastic Q-Soft-Bayes (70,227 seconds vs. 160,849 seconds), although their per-iteration time are the same in the big-O notation.
Recall that each iteration of Stochastic Q-Soft-Bayes involves two matrix logarithms and one matrix exponential; 
either is slower than the eigendecomposition needed in Algorithm~\ref{alg:1}. 
This explains why Algorithm~\ref{alg:1} is faster than stochastic Q-Soft-Bayes in practice. 

The two stochastic methods are still slower than $R \rho R$ and monotonous Frank-Wolfe in terms of the elapsed time, showing some space for improvement for stochastic methods. 
Notice that both the number of iterations and per-iteration time affect the real elapsed time. 
$R \rho R$ and monotonous Frank-Wolfe actually have significantly shorter per-iteration time; 
$O ( d^\omega )$ for the former and $O ( d^2 )$ for the latter. 
Nevertheless, it seems difficult to characterize the required number of iterations, because $R \rho R$ and monotonous Frank-Wolfe lack clear non-asymptotic convergence guarantees.

\acks{C.-E.~Tsai and Y.-H.~Li are supported by the Young Scholar Fellowship (Einstein Program) of the National Science  and Technology Council of Taiwan under grant numbers MOST 108-2636-E-002-014, MOST 109-2636-E-002-025, MOST 110-2636-E-002-012 and MOST 111-2636-E-002-019 and by the research project ``Pioneering Research in Forefront Quantum Computing, Learning and Engineering'' of National Taiwan University under grant number NTU-CC-111L894606. 

H.-C.~Cheng is supported by the Young Scholar Fellowship (Einstein Program) of the National Science  and Technology Council (NSTC) in Taiwan (R.O.C.) under Grant MOST 111-2636-E-002-026, Grand MOST 111-2119-M-007-006, Grant MOST 111-2119-M-001-004, and is supported by the Yushan Young Scholar Program of the Ministry of Education in Taiwan (R.O.C.) under Grant NTU-111V1904-3, and Grant NTU-111L3401, and by the research project ``Pioneering Research in Forefront Quantum Computing, Learning and Engineering'' of National Taiwan University under Grant No.~NTU-CC-111L894605."}

\bibliographystyle{unsrtnat}
\bibliography{omd_qst}

\begin{thebibliography}{35}
\providecommand{\natexlab}[1]{#1}
\providecommand{\url}[1]{\texttt{#1}}
\expandafter\ifx\csname urlstyle\endcsname\relax
  \providecommand{\doi}[1]{doi: #1}\else
  \providecommand{\doi}{doi: \begingroup \urlstyle{rm}\Url}\fi

\bibitem[Hradil(1997)]{Hradil1997aa}
Z.~Hradil.
\newblock Quantum-state estimation.
\newblock \emph{Phys. Rev. A}, 55, 1997.

\bibitem[Bolduc et~al.(2017)Bolduc, Knee, Gauger, and Leach]{Bolduc2017}
Eliot Bolduc, George~C. Knee, Erik~M. Gauger, and Jonathan Leach.
\newblock Projected gradient descent algorithms for quantum state tomography.
\newblock \emph{{npj} Quantum Inf.}, 3, 2017.

\bibitem[Kyrillidis et~al.(2013)Kyrillidis, Becker, Cevher, and
  Koch]{Kyrillidis2013}
Anastasios Kyrillidis, Stephen Becker, Volkan Cevher, and Christoph Koch.
\newblock Sparse projections onto the simplex.
\newblock In \emph{Proc. 30th Int. Conf. Machine Learning}, pages 235--243,
  2013.

\bibitem[Alman and Williams(2021)]{Alman2021}
Josh Alman and Vriginia~Vassilevska Williams.
\newblock A refined laser method and faster matrix multiplication.
\newblock In \emph{Proc. 2021 ACM-SIAM Symp. Discrete Algorithms (SODA)}, 2021.

\bibitem[Chen et~al.(2022)Chen, Huang, Li, Liu, and Sellke]{Chen2022aa}
Sitan Chen, Brice Huang, Jerry Li, Allen Liu, and Mark Sellke.
\newblock Tight {B}ounds for {S}tate {T}omography with {I}ncoherent
  {M}easurements.
\newblock 2022.
\newblock arXiv:2206.05265.

\bibitem[Nesterov(2018)]{Nesterov2018a}
Yurii Nesterov.
\newblock \emph{Lectures on Convex Optimization}.
\newblock Springer, Cham, CH, second edition, 2018.

\bibitem[Knee et~al.(2018)Knee, Bolduc, Leach, and Gauger]{Knee2018a}
George~C. Knee, Eliot Bolduc, Jonathan Leach, and Erik~M. Gauger.
\newblock Quantum process tomography via completely positive and
  trace-preserving projection.
\newblock \emph{Phys. Rev. A}, 2018.

\bibitem[You et~al.(2022)You, Cheng, and Li]{You2022}
Jun-Kai You, Hao-Chung Cheng, and Yen-Huan Li.
\newblock Minimizing quantum {R}{\'{e}}nyi divergences via mirror descent with
  {P}olyak step size.
\newblock In \emph{IEEE Int. Symp. Information Theory (ISIT)}, 2022.

\bibitem[Lin et~al.(2021)Lin, Hsu, and Li]{Lin2021b}
Chien-Ming Lin, Yu-Ming Hsu, and Yen-Huan Li.
\newblock An online algorithm for maximum-likelihood quantum state tomography.
\newblock 2021.
\newblock arXiv:2012.15498.

\bibitem[Nielsen and Chuang(2010)]{Nielsen2010}
Michael~A. Nielsen and Isaac~L. Chuang.
\newblock \emph{Quantum Computation and Quantum Information}.
\newblock Cambridge Univ. Press, Cambridge, UK, 2010.

\bibitem[Gross et~al.(2010)Gross, Liu, Flammia, Becker, and Eisert]{Gross2010}
David Gross, Yi-Kai Liu, Steven~T. Flammia, Stephen Becker, and Jens Eisert.
\newblock Quantum state tomography via compressed sensing.
\newblock \emph{Phys. Rev. Lett.}, 105, 2010.

\bibitem[Blume-Kohout(2010)]{Blume-Kohout2010}
Robin Blume-Kohout.
\newblock Hedged maximum likelihood quantum state estimation.
\newblock \emph{Phys. Rev. Lett.}, 105, 2010.

\bibitem[O'Donnell and Wright(2016)]{ODonnell2016}
Ryan O'Donnell and John Wright.
\newblock Efficient quantum tomography.
\newblock In \emph{Proc. 48th Annu. ACM Symp. Theory of Computing}, pages
  899--912, 2016.

\bibitem[Haah et~al.(2017)Haah, Harrow, Ji, Wu, and Yu]{Haah2017}
Jeongwan Haah, Aram~W. Harrow, Zhengfeng Ji, Xiaodi Wu, and Nengkun Yu.
\newblock Sample-optimal tomography of quantum states.
\newblock \emph{IEEE Trans. Inf. Theory}, 63\penalty0 (9):\penalty0 5628--5641,
  2017.

\bibitem[Torlai et~al.(2018)Torlai, Mazzola, Carrasquilla, Troyer, Melko, and
  Carleo]{Torlai2018}
Giacomo Torlai, Guglielmo Mazzola, Juan Carrasquilla, Matthias Troyer, Roger
  Melko, and Giuseppe Carleo.
\newblock Neural-network quantum state tomography.
\newblock \emph{Nat. Phys.}, 14:\penalty0 447--450, 2018.

\bibitem[Gu\c{t}\v{a} et~al.(2020)Gu\c{t}\v{a}, Kahn, Kueng, and
  Tropp]{Guta2020}
M.~Gu\c{t}\v{a}, J.~Kahn, R.~Kueng, and J.~A. Tropp.
\newblock Fast state tomography with optimal error bounds.
\newblock \emph{J. Phys. A: Math. Theor.}, 53, 2020.

\bibitem[Scholten and Blume-Kohout(2018)]{Scholten2018}
Travis~L. Scholten and Robin Blume-Kohout.
\newblock Behavior of maximum likelihood in quantum state tomography.
\newblock \emph{New J. Phys.}, 20, 2018.

\bibitem[{\v R}eh{\'a}{\v c}ek et~al.(2007){\v R}eh{\'a}{\v c}ek, Hradil,
  Knill, and Lvovsky]{Rehacek2007aa}
Jaroslav {\v R}eh{\'a}{\v c}ek, Zden{\v e}k Hradil, E.~Knill, and A.~I.
  Lvovsky.
\newblock Diluted maximum-likelihood algorithm for quantum tomography.
\newblock \emph{Phys. Rev. A}, 75\penalty0 (4):\penalty0 042108, 2007.

\bibitem[Tran-Dinh et~al.(2015)Tran-Dinh, Kyrillidis, and
  Cevher]{Tran-Dinh2015aa}
Quoc Tran-Dinh, Anastasios Kyrillidis, and Volkan Cevher.
\newblock Composite {S}elf-{C}oncordant {M}inimization.
\newblock \emph{Journal of Machine Learning Research}, 16\penalty0
  (12):\penalty0 371--416, 2015.

\bibitem[Bauschke et~al.(2017)Bauschke, Bolte, and Teboulle]{Bauschke2017}
Heinz~H. Bauschke, J\'{e}r\^{o}me Bolte, and Marc Teboulle.
\newblock A descent lemma beyond {L}ipschitz gradient continuity: first-order
  methods revisited and applications.
\newblock \emph{Math. Oper. Res.}, 42\penalty0 (2):\penalty0 330--348, 2017.

\bibitem[Li and Cevher(2019)]{Li2019a}
Yen-Huan Li and Volkan Cevher.
\newblock Convergence of the exponentiated gradient method with {A}rmijo line
  search.
\newblock \emph{J. Optim. Theory Appl.}, 181\penalty0 (2):\penalty0 588--607,
  2019.

\bibitem[Dvurechensky et~al.(2020)Dvurechensky, Ostroukhov, Safin, Shtern, and
  Staudigl]{Dvurechensky2020}
Pavel Dvurechensky, Petr Ostroukhov, Kamil Safin, Shimrit Shtern, and Mathias
  Staudigl.
\newblock Self-concordant analysis of {F}rank-{W}olfe algorithms.
\newblock In \emph{Proc. 37th Int. Conf. Machine Learning}, 2020.

\bibitem[Carderera et~al.(2021)Carderera, Besan\c{c}on, and
  Pokutta]{Carderera2021aa}
Alejandro Carderera, Mathieu Besan\c{c}on, and Sebastian Pokutta.
\newblock Simple steps are all you need: {F}rank-{W}olfe and generalized
  self-concordant functions.
\newblock In \emph{Adv. Neural Information Processing Systems 34}, volume~34,
  pages 5390--5401, 2021.

\bibitem[Zhao and Freund(2022)]{Zhao2022aa}
Renbo Zhao and Robert~M. Freund.
\newblock Analysis of the {F}rank--{W}olfe method for convex composite
  optimization involving a logarithmically-homogeneous barrier.
\newblock \emph{Math. Program.}, 2022.

\bibitem[Youssry et~al.(2019)Youssry, Ferrie, and Tomamichel]{Youssry2019aa}
Akram Youssry, Christopher Ferrie, and Marco Tomamichel.
\newblock Efficient online quantum state estimation using a
  matrix-exponentiated gradient method.
\newblock \emph{New J. of Phys.}, 21\penalty0 (3):\penalty0 033006, 2019.

\bibitem[Yang et~al.(2020)Yang, Jiang, Zhang, and Sun]{Yang2020}
Feidiao Yang, Jiaqing Jiang, Jialin Zhang, and Xiaoming Sun.
\newblock Revisiting online quantum state learning.
\newblock In \emph{Proc. AAAI Conf. Artificial Intelligence}, 2020.

\bibitem[Zimmert et~al.(2022)Zimmert, Agarwal, and Kale]{Zimmert2022}
Julian Zimmert, Naman Agarwal, and Satyen Kale.
\newblock Pushing the efficiency-regret {P}areto frontier for online learning
  of portfolios and quantum states.
\newblock In \emph{Proc. 35th Conf. Learning Theory}, pages 182--226, 2022.

\bibitem[Aaronson(2020)]{Aaronson2020}
Scott Aaronson.
\newblock Shadow tomography of quantum states.
\newblock \emph{SIAM J. Comput.}, 49\penalty0 (5):\penalty0
  STOC18--368--STOC18--394, 2020.

\bibitem[Aaronson et~al.(2018)Aaronson, Chen, Hazan, Kale, and
  Nayak]{Aaronson2018}
Scott Aaronson, Xinyi Chen, Elad Hazan, Satyen Kale, and Ashwin Nayak.
\newblock Online learning of quantum states.
\newblock In \emph{Adv. Neural Information Processing Systems 31}, 2018.

\bibitem[Cesa-Bianchi et~al.(2004)Cesa-Bianchi, Conconi, and
  Gentile]{Cesa-Bianchi2004aa}
N.~Cesa-Bianchi, A.~Conconi, and C.~Gentile.
\newblock On the generalization ability of on-line learning algorithms.
\newblock \emph{IEEE Trans. Inf. Theory}, 50\penalty0 (9):\penalty0 2050--2057,
  2004.

\bibitem[Cutkosky(2019)]{Cutkosky2019aa}
Ashok Cutkosky.
\newblock Anytime {O}nline-to-{B}atch, {O}ptimism and {A}cceleration.
\newblock In \emph{Proc. 36th Int. Conf. on Mach. Learn.}, volume~97, pages
  1446--1454, 2019.

\bibitem[Kot{\l}owski and Neu(2019)]{Kotlowski2019}
Wojciech Kot{\l}owski and Gergely Neu.
\newblock Bandit principal component analysis.
\newblock In \emph{Proc. 32nd Conf. Learning Theory}, pages 1994--2024, 2019.

\bibitem[Tsai et~al.(2022)Tsai, Cheng, and Li]{Tsai2022aa}
Chung-En Tsai, Hao-Chung Cheng, and Yen-Huan Li.
\newblock Online {S}elf-{C}oncordant and {R}elatively {S}mooth {M}inimization,
  {W}ith {A}pplications to {O}nline {P}ortfolio {S}election and {L}earning
  {Q}uantum {S}tates.
\newblock 2022.
\newblock arXiv:2210:00997 [stat.ML].

\bibitem[Lvovsky(2004)]{Lvovsky2004aa}
A.~I. Lvovsky.
\newblock Iterative maximum-likelihood reconstruction in quantum homodyne
  tomography.
\newblock \emph{J. of Opt. B: Quantum and Semiclass. Opt.}, 6\penalty0
  (6):\penalty0 S556, 2004.

\bibitem[Molina-Terriza et~al.(2004)Molina-Terriza, Vaziri,
  \v{R}eh\'{a}\v{c}ek, Hradil, and Zeilinger]{MolinaTerriza2004}
G.~Molina-Terriza, A.~Vaziri, J.~\v{R}eh\'{a}\v{c}ek, Z.~Hradil, and
  A.~Zeilinger.
\newblock Triggered qutrits for quantum communication protocols.
\newblock \emph{Phys. Rev. Lett.}, 92\penalty0 (16), 2004.

\end{thebibliography}

\end{document}